\documentclass{aa}                    
\usepackage{graphicx}                
\usepackage{txfonts}                 

\newcommand{\kms}{km\,s$^{-1}$}
\newcommand{\ms}{m\,s$^{-1}$}

\begin{document}

\title{PEPSI deep spectra\thanks{Based on data acquired with PEPSI fed by the solar disk integration (SDI) telescope operated by AIP at the Large Binocular Telescope Observatory (LBTO). The LBT is an international collaboration among institutions in the United States, Italy and Germany. LBT Corporation partners are: The University of Arizona on behalf of the Arizona Board of Regents; Istituto Nazionale di Astrofisica, Italy; LBT Beteiligungsgesellschaft, Germany, representing the Max-Planck Society, The Leibniz Institute for Astrophysics Potsdam (AIP), and Heidelberg University; The Ohio State University, and The Research Corporation, on behalf of The University of Notre Dame, University of Minnesota and University of Virginia.}}

\subtitle{I. The Sun-as-a-star}

\author{K. G. Strassmeier, I. Ilyin, \and M. Steffen}

\institute{Leibniz-Institute for Astrophysics Potsdam (AIP), An der Sternwarte 16, D-14482 Potsdam, Germany; \\ \email{kstrassmeier@aip.de}, \email{ilyin@aip.de}, \email{msteffen@aip.de}}

\date{Received ... ; accepted ...}

\abstract
{Full-disk solar flux spectra can be directly compared to stellar spectra and thereby serve as our most important reference source for, for example stellar chemical abundances, magnetic activity phenomena, radial-velocity signatures or global pulsations.}
{As part of the first Potsdam Echelle Polarimetric and Spectroscopic Instrument (PEPSI) key-science project, we aim to provide well-exposed and average-combined  ({\sl viz.} deep) high-resolution spectra of representative stellar targets. Such deep spectra contain an overwhelming amount of information, typically much more than what could be analyzed and discussed within a single publication.  Therefore, these spectra will be made available in form of (electronic) atlases. The first star in this series of papers is our Sun. It also acts as a system-performance cornerstone.}
{The Sun was monitored with PEPSI at the Large Binocular Telescope (LBT). Instead of the LBT we used a small robotic solar disk integration (SDI) telescope. The deep spectra in this paper are the results of combining up to $\approx$100 consecutive exposures per wavelength setting and are compared with other solar flux atlases.  }
{Our software for the optimal data extraction and reduction of PEPSI spectra is described and verified with the solar data. Three deep solar flux spectra with a spectral resolution of up to 270,000, a continuous wavelength coverage from 383\,nm to 914\,nm, and a photon signal to noise ratio (S/N) of between 2,000-8,000:1 depending on wavelength are presented. Additionally, a time-series of 996 high-cadence spectra in one cross disperser is used to search for intrinsic solar modulations. The wavelength calibration based on Th-Ar exposures and simultaneous Fabry-P\'erot combs enables an absolute wavelength solution within 10~\ms\ (rms) with respect to the HARPS laser-comb solar atlas and a relative rms of 1.2\,\ms\ for one day. For science demonstration, we redetermined the disk-average solar Li abundance to  1.09$\pm$0.04\,dex on the basis of 3D NLTE model atmospheres. We detected disk-averaged p-mode RV oscillations with a full amplitude of 47\,cm\,s$^{-1}$ at 5.5\,min. }
{Comparisons with two solar FTS atlases, as well as with the HARPS solar atlas, validate the PEPSI data product.  Now, PEPSI/SDI solar-flux spectra are being taken with a sampling of one deep spectrum per day, and are supposed to continue a full magnetic cycle of the Sun.}



\titlerunning{PEPSI deep spectra. I. The Sun as a star}

\maketitle

\section{Introduction}

Cool stars similar to the Sun exhibit such a rich amount of spectral lines that high spectral resolution is mandatory to see their subtle details and arrive at a correct interpretation. A good example is the problem of the solar lithium abundance (e.g., Brault \& M\"uller \cite{bra:mue}), which must be derived from a spectral line of only a few m\AA\ of equivalent width. Its value is particularly important because of its reference for stellar nucleosynthesis and evolution models. Compared to the meteoritic Li abundance of 3.28$\pm$0.05 (Lodders \cite{lodd}), lithium is depleted by a factor 150 in the solar photosphere. This amount of depletion cannot be explained by standard solar evolution models but requires additional mixing beyond the base of the convection zone.

Besides the Li problem, a multitude of other science cases has led in the past to the creation of high-fidelity solar spectral atlases either disk centered or disk averaged. Among the first was the Utrecht atlas by Minnaert et al. (\cite{utrecht}) and several followed, most notably the Sacramento Peak atlas (Beckers et al. \cite{beckers}), the Li\`ege atlas (Delbouille et al. \cite{liege}) and the Hamburg atlas (Neckel \& Labs \cite{nek:lab}) (see the recent comparison of the latter two by Doerr et al. \cite{doerr}). Subsequent solar atlases were also used to identify blended transitions in stellar spectra that otherwise do not reach comparable spectral resolutions. Also, the Sun is a long-period variable with a cyclic changing ``solar constant'' (c.f. Fr\"ohlich \cite{fro}). Monitoring it on a synoptic basis proved extremely valuable (Livingston et al.~\cite{liv}). Among Livingston et al.'s findings was that not only Ca\,{\sc ii} H\&K traced the magnetic cycle of the Sun but that even the cores of high photosphere lines likely vary with the 22-yr Hale cycle. The latter has yet to be verified and understood. With PEPSI and its solar feed we are now able to observe the Sun on a daily basis and thus may verify Livingston et al.'s claim. In any case, solar feeds for modern high-precision (night-time) spectrographs became more valued in the past due to the ever increasing importance of the solar-stellar connection in the exoplanet era (e.g., Pevtsov et al.~\cite{pev:ber}, Hall \& Lockwood \cite{sss}, Lanza et al.~\cite{lanza}). In addition, solar data allow a comparison of properties of disk-resolved images with disk-integrated spectra, which is useful for the interpretation of stellar Doppler images (Strassmeier \cite{spots}).

\begin{figure*}
\includegraphics[angle=0,width=\textwidth]{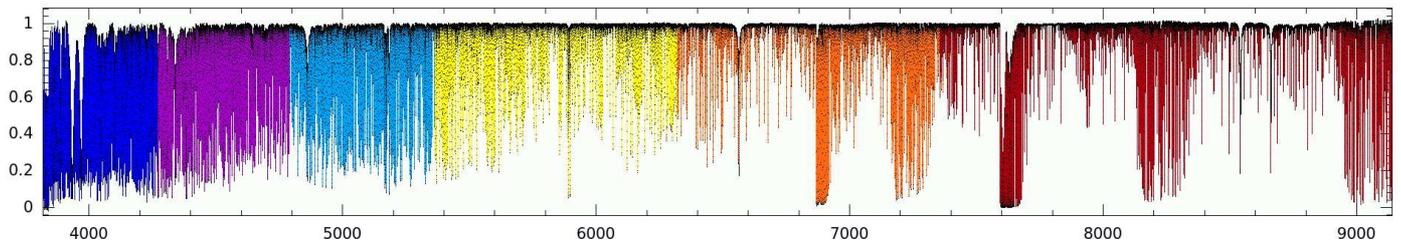}
\caption{Representative PEPSI solar spectrum taken with the SDI telescope (x-axis is wavelength in Angstroems, y-axis is relative intensity). Each spectrum covers the wavelength range 382.2--914.6\,nm without gaps with a sampling of 430,000 pixels at an average dispersion of 12~m\AA/pix. The different colors indicate the wavelength regions covered by the six cross dispersers.}
 \label{F1}
\end{figure*}

\begin{table*}
\caption{Summary of PEPSI deep solar spectra. } \label{T1}
\begin{tabular}{lllllllll}
\hline\hline \noalign{\smallskip}
Day (UT)            & I/404$^d$ & II/450 & III/508 & IV/584 & V/685 & VI/825 \\
\noalign{\smallskip}\hline \noalign{\smallskip}
$R = \lambda/\Delta\lambda$$^a$ & 220,000 & 230,000 & 240,000 & 250,000 & 240,000 & 240,000 \\
S/N$^b$ single spectrum         & 330 & 350 & 370 & 620 & 960 & 590 \\
Exp. time single spectrum (s)   & 15 & 5 & 3 & 2 & 2 & 1.5 \\
\noalign{\smallskip}\hline \noalign{\bigskip}
$N$ single spectra$^c$   &  &  &  &  &  &  \\
\noalign{\smallskip}\hline \noalign{\smallskip}
Nov. 15, 2016   & 28 & 40 & 40 & 40 & 40 & 29 \\
Nov. 16, 2016   & 40 & 52 & 60 & 53 & 60 & 41 \\
Nov. 17, 2016   & 100 & 100 & 112 & 100 & 112 & 100 \\
\noalign{\smallskip}\hline\noalign{\bigskip}
S/N deep spectra   &  &  &  &  &  &  \\
\noalign{\smallskip}\hline \noalign{\smallskip}
Nov. 15, 2016   & 1050 & 2050 & 2850 & 3950 & 5150 & 3500 \\
Nov. 16, 2016   & 1250 & 2350 & 3450 & 4550 & 6350 & 4150 \\
Nov. 17, 2016   & 2000 & 3300 & 4700 & 6250 & 8600 & 6500 \\
\noalign{\smallskip}\hline\noalign{\bigskip}
$N$ for time series  &  &  &  &  &  &  \\
\noalign{\smallskip}\hline \noalign{\smallskip}
Oct. 14, 2016  & \dots & \dots & 263 & \dots & \dots & \dots \\
Oct. 15, 2016  & \dots & \dots & 386 & \dots & \dots & \dots \\
Oct. 16, 2016  & \dots & \dots & 347 & \dots & \dots & \dots \\
\noalign{\smallskip}\hline
\end{tabular}
\tablefoot{$^a$Median resolution at mid of \'echelle order. $^b$S/N per pixel in the continuum at the mid wavelength of each cross disperser. $^c$Number $N$ of individual spectra observed with PEPSI. $^d$I/404 (e.g.,) means CD\,I with center wavelength 404\,nm. The time series Oct.\,14-16, 2016 was taken only in CD\,III.}
\end{table*}

The first modern solar flux atlases were probably those of Kurucz et al. (\cite{kur:fur}) and Brault \& Neckel (\cite{bra:nec}), obtained with the McMath-Pierce telescope of the National Solar Observatory (NSO) at Kitt Peak and NSO's Fourier Transform Spectrograph (FTS). These were succeeded by the atlases of Wallace et al. (\cite{nso-atlas}) for the disk center and Wallace et al. (\cite{nso}) for the integrated disk, obtained with the same telescope and FTS but corrected for telluric features and with an empirical Doppler correction. Both atlases reached a resolution between 350,000 and 700,000 depending on wavelength. Extensions to the infrared (0.7 to 22\,$\mu$m) were added in separate volumes  (e.g., Wallace et al. \cite{nso-ir}). Later, Kurucz (\cite{kur}) reprocessed the old 1980-81 data and incorporated a synthetic telluric spectrum to model the telluric contamination. More recently, Reiners et al. (\cite{iag}) introduced a new FTS solar flux atlas obtained with a small telescope and a Bruker FTS at the Institute for Astrophysics in G\"ottingen (IAG) for the wavelength range 405--2300\,nm. From a comparison with the Wallace et al. Kitt Peak NSO atlases the authors arrived at a  wavelength calibration for the IAG atlas accurate to $\pm$10~\ms\ for the 405--1065\,nm region, while the Kitt Peak atlases showed deviations between several ten to 100~\ms .

\begin{figure*}
\includegraphics[angle=0,width=\textwidth,clip]{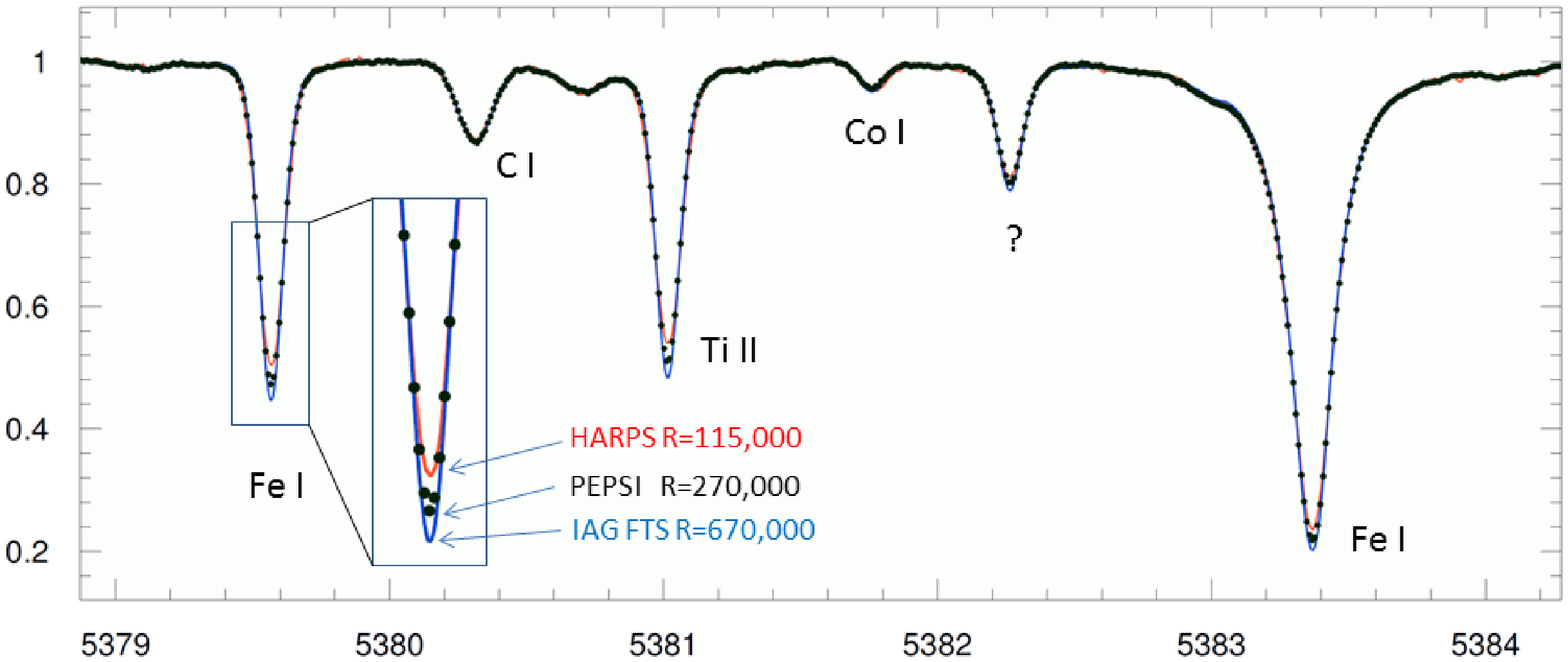}
\caption{Wavelength region around the solar C\,{\sc i} line at 5380.3\,\AA . Dots indicate one of the PEPSI deep spectra and the dark blue and bright red line is the IAG FTS atlas from Reiners et al. (\cite{iag}) and the HARPS spectrum from Molaro et al. (\cite{molaro}), respectively. The zoom shows the Fe\,{\sc i} 5379.6-\AA\ line. The FTS atlas has a spectral resolution of $\approx$670,000, the HARPS spectrum 115,000. }
 \label{F2}
\end{figure*}

For stars, radial velocity (RV) stability became center stage after the first extra-solar planets were discovered in the nineties and a very accurate wavelength calibration became a necessity. The spearhead instrument in this respect was ESO's HARPS spectrograph (Pepe et al.~\cite{harps}) which is now equipped, like its northern-hemisphere counterpart on La Palma (HARPS-N; Cosentino et al. \cite{harps-n}), with a fibre-based laser frequency comb for wavelength calibration. Laser-comb calibrated solar spectroscopy was investigated previously by Probst et al. (\cite{probst}). Solar-flux spectra were obtained with both HARPS instruments and an $R\approx$115,000 comb-calibrated atlas for two wavelength ranges between 476 and 585\,nm was presented by Molaro et al. (\cite{molaro}). Only in this work it was realized that there is a wavelength distortion in every HARPS \'echelle order of up to $\pm$40\,\ms\ caused by the stitching of CCD sections. Sun-as-a-star observations with HARPS-N by Dumusque et al. (\cite{dum}) led to a time series with residual RVs of 60\,cm\,s$^{-1}$ after correcting for sunspot and plage perturbations (see also Haywood et al. \cite{hay:acc}, Molaro \& Munai \cite{mol:mon} and Molaro et al. \cite{mol:lan} for asteroid-based solar-RV observations). While the RV stability of HARPS remained unaffected, it nevertheless proved how important it is to use the Sun for a detailed spectrograph characterization.

We have just put into operation the optical high-resolution \'echelle spectrograph PEPSI (Strassmeier et al. \cite{pepsi}) at the effective 11.8\,m Large Binocular Telescope (LBT; Hill et al. \cite{lbt}). PEPSI provides a nominal spectral resolution of up to 270,000 for the wavelength range 383--907\,nm and can alternatively be (fibre) fed by the nearby 1.8\,m Vatican Advanced Technology Telescope (VATT) of the Vatican Observatory or by a 13\,mm Solar Disk Integration (SDI) telescope. We emphasize that the   VATT and the SDI feed starlight or Sunlight in exactly the same way to the spectrograph as during nominal LBT use which allows for a straightforward comparison of solar and stellar spectra. PEPSI and SDI take at least one, typically three, ``deep'' solar spectra on a daily basis. The aim is to cover the full optical spectrum several times a day for the entirety of the solar magnetic activity cycle, continuing the program of Livingston et al. (\cite{liv}) but with extended spectral coverage and enhanced cadence. A similar attempt is already ongoing at NSO with its SOLIS/ISS project (Pevtsov et al.~\cite{pev:ber}) in a number of selected wavelength regions but with an instrument not applicable to night-time stellar observations. SOLIS's integrated sunlight spectrometer (ISS) is a double-pass spectrograph that produces daily high-quality spectra for Sun-as-a-star studies in eight typically 1-nm wide wavelength windows (Keller et al. \cite{iss}).

In the current paper, we present first PEPSI spectra of the Sun observed as a star.  These initial data were taken during October and November 2016 while still in commissioning. Deep spectra of the {\sl Gaia} benchmark stars and other M-K standards will be presented in paper~II (Strassmeier et al.~\cite{pap2}) and a chemical analysis of the ancient planet-system host star Kepler-444 is presented in paper~III (Mack et al.~\cite{pap3}). Section~\ref{S2} describes the observations and the spectrograph while Sect.~\ref{S3} gives details of the data extraction and reduction. In Section~\ref{S4}, we compare our spectra with other solar-flux spectra and describe its characteristics in terms of S/N, resolution, wavelength scale and various blemishes.  Section~\ref{S5} presents our detection of the solar 5-min oscillation and an attempt to redetermine the solar Li abundance from deep spectra with the most up-to-date non-LTE 3D model atmospheres and line blending. Section~\ref{S6} is a summary.

\section{Observations}\label{S2}

Solar flux spectra were obtained with the Potsdam Echelle Polarimetric and Spectroscopic Instrument (Strassmeier et al. \cite{pepsi}) and the 13\,mm diameter SDI telescope. PEPSI is an asymmetric fiber-fed white-pupil \'echelle spectrograph with two arms (blue and red optimized) covering the wavelength range 383--914\,nm with six different cross dispersers (CD). In this paper the wavelength coverage was slightly extended on both wavelength edges by extracting an extra half of an \'echelle order. The instrument is stabilized in a pressure and thermally controlled chamber and is fed by three pairs of octagonal fibres per telescope. The different core diameters of the fibres and their respective image slicers set the three different nominal resolutions of the spectrograph (43,000, 120,000 and 250,000). The 250,000-mode is used for the spectra in this paper and is made possible with a 7-slice image slicer and a 100-$\mu$m fibre core through a projected sky aperture of 0.74\arcsec , comparable to the median seeing of the LBT site. Its resolution element is sampled with two pixels. All observations in this paper were made with the version-2 waveguide image slicer described in Beckert et al. (\cite{beck:is2}). Two 10.3k$\times$10.3k STA1600LN (Bredthauer et al.~\cite{1600LN}) CCDs with 9-$\mu$m pixels record a total of 92 \'echelle orders in the six wavelength settings. The average dispersion is 7~m\AA/pix.

The SDI telescope is mounted on the kitchen balcony of the LBT-Observatory and consists of an effective 13\,mm diameter reflecting telescope, with a $f=100$\,mm off-axis-parabola illuminating an integration sphere. The integration sphere feeds light into a pair of fibres, just like for the LBT. The SDI telescope itself is fully robotic. It acquires the Sun shortly after sunrise and guides on it all day long until sunset. Daily monitoring of the Sun in the full wavelength range of PEPSI is SDI's default science operation. During November 15-17, 2016, daily consecutive exposures were taken with integration times of 15\,s, 5\,s, 3\,s, 2\,s, 2\,s, and 1.5\,s for the six wavelength settings from blue to red, respectively.

The cadence is driven by the CCD overhead rather than the exposure time; 55\,s CCD readout, 9\,s transfer and disk writing, 4\,s CCD cleaning and 2\,s other overhead, in total 70\,s. Then, the entire cycle with all wavelength settings is repeated 3--4 times depending on weather conditions. Individual exposures reach signal-to-noise ratios (S/N) between 50:1 in the cores of the Ca\,{\sc ii} H\&K lines in 15\,s and about 850:1 for continuum regions near 700\,nm in 1.5\,s. One full deep spectrum consists of a total of between 28-112 individual exposures per CD and peaks at a formal S/N of 8,600:1. We thus record up to $\approx$600 solar exposures per day. During October 14-16, 2016, a time series of 996 consecutive spectra was taken in CD\,III. Its log is given in Table~\ref{T1}. The upper part of Table~\ref{T1} is the observing log for the deep spectra from November, while the lower part is the log of the time-series data from October.

The six wavelength settings are defined by the six cross dispersers, CD\,I to VI, three per arm and two always simultaneously. Thus, it takes three exposures to cover the entire wavelength range as shown in Fig.~\ref{F1}. Due to the binocular nature of the LBT (and the SDI), each of the two CCDs record two target exposures for every \'echelle order simultaneously, one from each of the binocular telescopes (called SX and DX). Additionally, these two target spectra are flanked by one ``sky'' fibre each. The time-series SDI observations in October 2016 used both available sky fibres for simultaneous recording of the light from a stabilized Fabry-P\'erot etalon (FPE). These exposures consist thus of four spectra per \'echelle order in the spatial sequence FPE-target-target-FPE, that is, wavelength calibration is recorded on the same exposure and spatially adjacent to the target spectra. Thousands of flat field exposures from a halogen lamp and bias readouts were recorded throughout the observing program and are used to create the super-master flat (same lamp for all four simultaneous spectra).

Reduced FITS files of the three deep solar spectra in Table~\ref{T1} are made available just like a solar atlas and are deposited for download (see Sect.~\ref{S6}). The time-series data are also available but on individual demand. The FITS headers contain all necessary meta data. All spectra are shifted to solar barycentric velocity based on the KPNO Th-Ar wavelengths.

\section{Details of the CCD data reduction}\label{S3}

The Spectroscopic Data Systems for PEPSI (SDS4PEPSI) is a generic package in C++ under Linux based on a numerical template library and graphical toolkits. It is implemented as the control system for various distributed units of the PEPSI spectrograph and is also used for comprehensive \'echelle image processing, reduction, and analysis of the resulting spectra. As such, it is based on the 4A package for the NOT spectrograph SoFin (Ilyin \cite{4A}). We apply SDS4PEPSI the same way for regular stellar spectra or for the present solar spectra. The incorporated image processing and reduction toolkit relies on adaptive selection of parameters by using statistical inference and robust estimators. The standard components include bias overscan detection and subtraction, scattered light surface extraction and subtraction, definition of \'echelle orders, weighted extraction of spectral orders, wavelength calibration, and a self-consistent continuum fit to the full 2D image of extracted orders. The final data product is shown in Fig.~\ref{F2} for a 6-\AA\ wavelength excerpt of a deep solar spectrum around the carbon C\,{\sc i} line at 5380.3\,\AA . Also shown in this figure is a comparison with the IAG FTS and the HARPS atlases. The line depths due to the three different spectral resolutions are highlighted.

\begin{figure*}
{\bf a.}\\
\includegraphics[angle=0,width=\textwidth, clip]{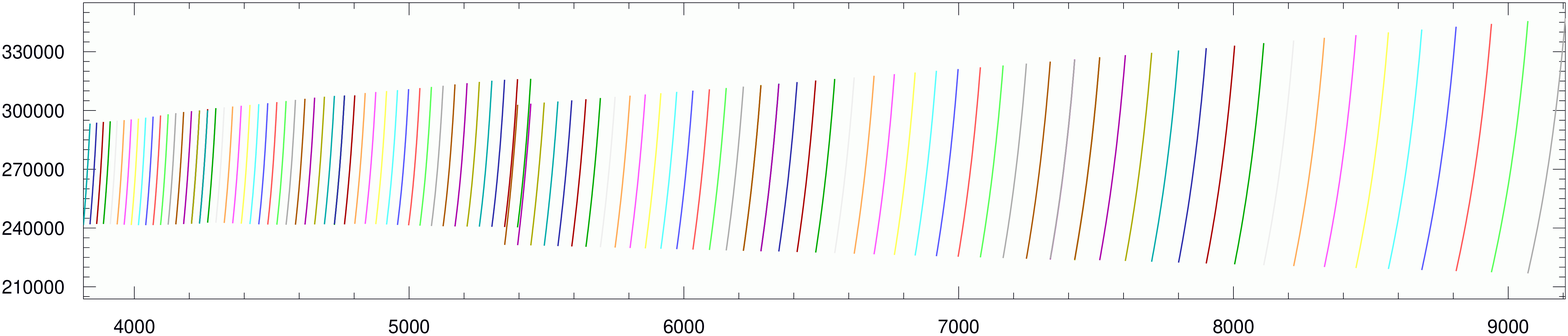}
{\bf b.}\\

\includegraphics[angle=0,width=\textwidth, clip]{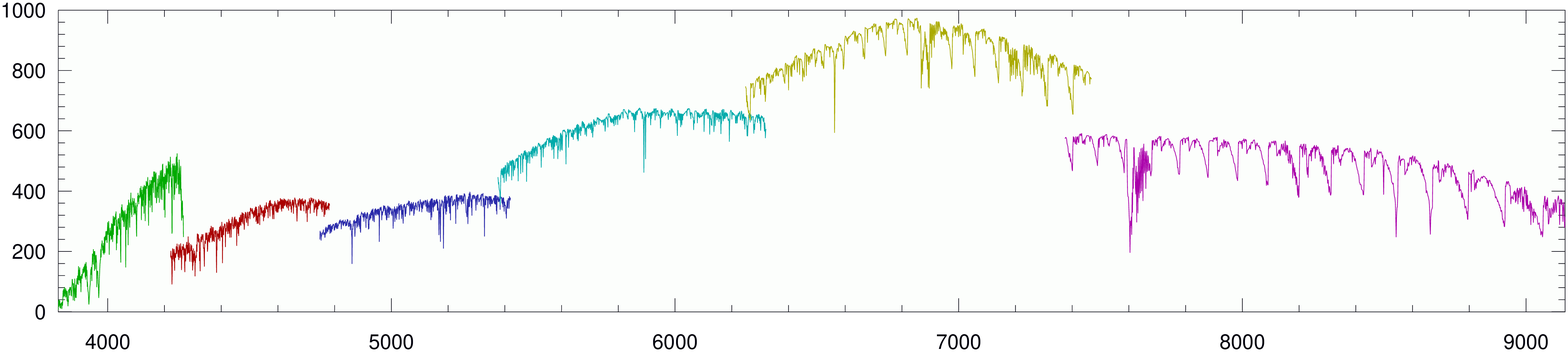}
\caption{PEPSI deep spectra characteristics. \emph{a.} Two-pixel spectrograph resolution as a function of wavelength. Shown is the free spectral range per \'echelle order. \emph{b.} S/N for an example spectrum. The spectra from the individual cross dispersers are identified with different colors and are, from left to right, CD\,I to CD\,VI.  Quasiperiodic peaks indicate the \'echelle orders. }
 \label{F3}
\end{figure*}

\subsection{CCD bias correction}

The CCD overscan in both directions is used for the bias level curves. The curves are smoothed with an appropriate spline function before subtracting. During the subtraction, the original FITS image is transformed from the original integer format to a floating-point image.

\subsection{Photon-noise estimation}

Once the bias is correctly removed, the photon noise is estimated for each CCD pixel and stored in the variance plane of the FITS image. The final variance of the reduced spectra defines the resulting S/N value for each wavelength pixel and results from the error propagation introduced at each step of the data-reduction process. The variance of pixel intensities is estimated according to the known read-out noise and the CCD gain factor.

The STA1600 CCD gain calibration is done for each of its 16 amplifiers by using the ratio of two de-focused images of master flat fields repeated at different illumination levels. The conversion factor is the intercept of the fit line to the gain factor as a function of ADUs. The variance for each pixel is estimated according to the linear fit. Instrumental details are given in Strassmeier et al. (\cite{pepsi}).

\subsection{Flat-field correction}\label{S3.3}

The flat-field correction removes the fixed pattern noise of the CCD. The flat-field spectrum must have sufficiently high S/N in order to avoid degradation of the scientific image, which is a particular challenge for solar-flux spectra. To achieve this, we employ master flat-field spectra with typical S/N ratios of 2,000:1. Division by the master flat field will still contribute to the variances of the object image according to the respective variances of the flat field pixels. It is normally expected that only one set of master flat fields for each spectral region, that is CD setting, is needed. Because the CCD pixel-to-pixel noise is different for each wavelength, it is required that the master flats are taken for each cross-disperser separately. Furthermore, each pixel shows its own non-linear response to different illumination levels. Therefore, we apply an illumination level correction in order to correct for the CCD spatial noise pattern for each pixel (see later in Sect.~\ref{S4.4}).

The master flat field images are obtained by de-focusing the halogen \'echelle image. The result is that the spectral orders are smeared out and no voids between spectral orders are left. One master flat field consists of a sum of 70  individual exposures. In order to account for the change of the response function versus illumination light level, we make such master flat fields also at 70 different exposure levels. The resulting super master flat field is the polynomial approximation of 7th degree of the response function versus number of photoelectrons for each CCD pixel. The normalized response function for each master flat field is obtained with a 2D smoothing spline fit which removes any gradients in the image. It preserves the structure of the spatial profile of the corrected image.

To reduce the effective gate resistance of the STA1600 CCDs the producer placed a metal grid over the CCD imaging area during production. Three micron metal lines run vertically and horizontally over the poly-silicon gates. This CCD bussing re-appears in our images as a fixed pattern and is mostly pronounced at red wavelengths where it is seen every 165 pixel rows with an amplitude of up to 5\%. A similar periodic structure exists along the CCD columns every 330 pixels. Another fixed-pattern contamination is seen with a $\sim$38 pixel periodicity and 1\,\%\ amplitude at very high exposure levels. It corresponds to pixel aliasing related to the contact points during the mixed mask generation of the CCD's poly-silicon gate structure.

\subsection{Scattered light subtraction}

Scattered light on the CCD image is formed by the many stray-light possibilities in an \'echelle spectrograph and the natural inter-order scattering. Scattered light acts like an extra continuum and results in a reduction of the residual depth of the spectral lines. The scattered light surface for our \'echelle images is reconstructed from the 2D least-squares fit of a constrained smoothing spline to the inter-order gaps (the constrained smoothing is necessary in order to prevent negative intensities). After the first initial fit of the spline to the image, the residuals of the data are analyzed with ordered statistics to mask out any echelle orders in the data. This is subsequently used for the fit in the next iteration. This process is applied to every relevant CCD image because the scattered-light contamination increases with the brightness of the target and will vary, for example when optics were re-coated or extra light traps installed. Solar spectra are naturally the most prone to scattered light. The following constant scattered-light contamination is seen in the spectra in this paper; up to 4\% in CD\,I, 2\% in CD\,II, 1\% in CD\,III, 2\% in CD\,IV, 1\% in CD\,V, and 0.5\% in CD\,VI, and was properly removed.

\subsection{Definition of spectral orders}

The order definition is obtained from the traced flat-field exposures made separately for each image slicer. The overall curvature and its sign is obtained from the global cross-correlation of all orders. An elongated Gaussian profile is matched to each slice of every order to form a 3D matrix of Chebyshev polynomials coefficients for the final global fit. The overall curvature and width of the traces are obtained from a global cross-correlation of adjacent CCD rows in dispersion direction. This first approximation is used to detect all spectral orders and their slices for the subsequent non-linear least-squares fit to its elongated Gaussian profiles. Note that, e.g., IRAF cannot deal with tilted image slicers. The final global approximation with a 3D Chebyshev polynomial involves the fit of each spectral order and their slices versus CCD row number. The polynomial degree in cross-dispersion direction is not fixed but is selected according to the curvature of the \'echelle orders. Image slices within \'echelle orders are well spaced and approximated by the first polynomial degree.

\begin{figure*}
\center
\includegraphics[angle=0,width=160mm]{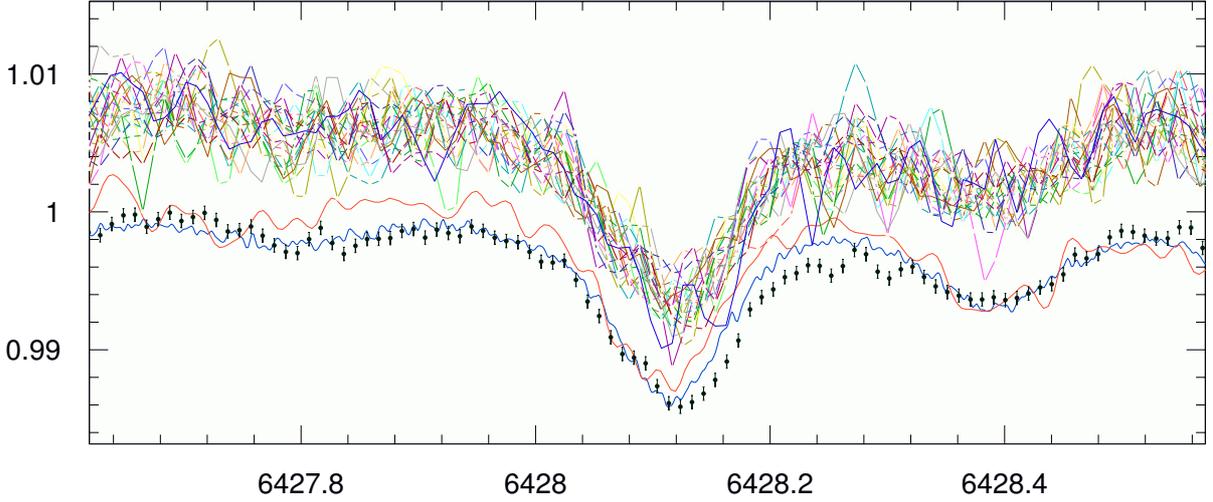}
\caption{Deep PEPSI spectrum (dots with error bars) and its individual 20 exposures (lines) for a 1-\AA\ wavelength region around the Cr\,{\sc i} 6428.1-\AA\ line. The twenty individual back-to-back spectra are offset in continuum by +0.01 for better visibility. The deep spectrum is unshifted and directly compared with the IAG (blue line) and NSO (red line) FTS atlases. We note that the equivalent width of the Cr line at 6428.1\,\AA\ is just 1.8~m\AA.}
 \label{F-6428}
\end{figure*}

\subsection{Optimal extraction of spectral orders}

Optimal extraction aims for a maximization of the S/N and for the rejection of cosmic-ray hits. The most important part of the optimal extraction is the calculation of the correct spatial profile for the full \'echelle image. The spatial profile, or illumination function, is derived from the raw image for every CCD pixel in each spectral order after normalization to the total flux. A robust spline fit is used to smooth the spatial profile along the dispersion direction. This is done for all orders in a global fit with a number of iterations in order to eliminate outliers in the data. The flux in every CCD pixel is formed as a weighted average with variances for all slices after the wavelength calibration. The cosmic-ray spikes are rejected during the fit according to a statistical test based on the pixel variances.

\subsection{Wavelength calibration}

The wavelength calibration is done for each image slicer separately by forming a 3D Chebyshev polynomial fit to its normalized wavelength, order number, and slice number. A typical error of the fit in the central part of a single CCD image is 3--5 \ms. The initial starting point for each image is given by the approximate value of the wavelength in the first \'echelle order of the spectrograph and the central order number for a given CD. With the use of a robust polynomial fit, it finds the best match of spectral-line wavelengths to the wavelength table.

The line-centering algorithm uses an un-weighted Gaussian fit to each emission line in each extracted slice in each spectral order. The centering algorithm usually selects about 70 of the best-fit emission lines. The wavelength calibration comprises a robust 3D Chebyshev polynomial fit to the wavelength table of the order number, slice number and CCD row. Robust means that deviating points in the wavelength fit are rejected with the use of an F-test. The wavelength identification is done with the aid of a search algorithm which gives a list of solutions with the best fit atop of it. The Th-Ar line list is a compilation from line lists of Norlen (\cite{nor}) (for Ar) and Palmer \& Engleman (\cite{pal:eng}) (for Th).

While the Th-Ar solution basically sets the absolute zero point of the wavelength calibration, the FPE sets the wavelength calibration from frame to frame. It also records the RV distortions due to, e.g., temperature variations in the spectrograph chamber e.g. introduced by coolant variations. In a recent paper, Bauer et al. (\cite{bauer}) studied the combined RV effects of hollow-cathode lamps with Fabry-P\'erot interferometers for HARPS and concluded that FPEs can reach a precision of 10~\ms\ when compared to a laser frequency comb.

Following standard practice the wavelengths are in air at standard temperature and pressure
(15$^\circ$C, 1013.25\,hPa; Edl\'en \cite{edlen}). A solar gravitational redshift of 633.4\,\ms\ was removed from the data. The wavelength scale of the spectra are reduced to the solar barycentric rest frame with the orbital and rotational velocities of the Earth removed, as well as the barycentric motion of the solar system due to rotation of the major planets. We use NOVAS\,2 to perform these velocity corrections (Kaplan et al. \cite{kaplan}).

\subsection{Shaping the spectra}

In this step, the orders are corrected for the \'echelle blaze function and also for vignetting effects if present. Another correction is due to the optical interference in the CCD substrate known as fringing. Its amplitude increases to a peak of 15\%\ but affects only the wavelengths of the reddest CD\,VI. All these effects are removed with a master flat field spectrum which is the sum of 300 individual exposures. The resulting S/N in such a spectrum is about 15,000 to ensure that there is no degradation after division of the solar spectrum. We found no change of the fringing amplitude versus light accumulation level in the flat field spectrum. The master flat field spectrum is extracted exactly the same way as any science image and used for normalization at the very last stage of image processing prior to continuum normalization. For the spectra in this paper, we were not able to distinguish between blaze and vignetting effects in wavelength direction which we interpret to be a sign of negligible vignetting if any. However, vignetting by the red-arm shutter at its upper edge was suspected but has been removed in the meantime.

\subsection{Global continuum fit and order merging}

For the solar spectra in this paper, we make use of the telluric-corrected FTS atlas of Wallace et al. (\cite{nso}) as a continuum guide. This is particular important for wavelengths shorter than 400\,nm, i.e. CD\,I, where the standard clip-and-fit algorithm fails because of the unspecified location of the continuum due to line blanketing.

For the rest of the spectrum and for stellar spectra, the continuum fit to the extracted spectral orders is done by default with a robust 2D smoothing spline on a regular grid of CCD pixels and \'echelle order numbers. The fit is made first in the cross-dispersion direction then in the dispersion direction. The essential part is to secure the spline from being affected by numerous outliers in form of absorption and emission lines. The outliers are detected and masked with the aid of robust noise estimators for which we use ordered statistics. A number of subsequent clip-and-fit iterations follow until the process is converged. The resulting regular and smoothed continuum surface provides the transition from the end of one spectral order to the beginning of the next.

\subsection{Telluric lines}

The blue part of the spectrum up to $\approx$500\,nm is essentially free of telluric lines. H$_2$O bands start to show up at this wavelength in the solar spectrum and by 579\,nm are joined by O$_2$ (see Wallace et al. \cite{nso}). While the O$_2$ bands are constant because well mixed in the atmosphere, the H$_2$O bands are highly variable and much more problematic. Our solar spectra are all contaminated by such lines. We did not attempt to correct for any of these telluric lines.

\begin{figure}
{\bf a.}\\
\includegraphics[angle=0,width=87mm,clip]{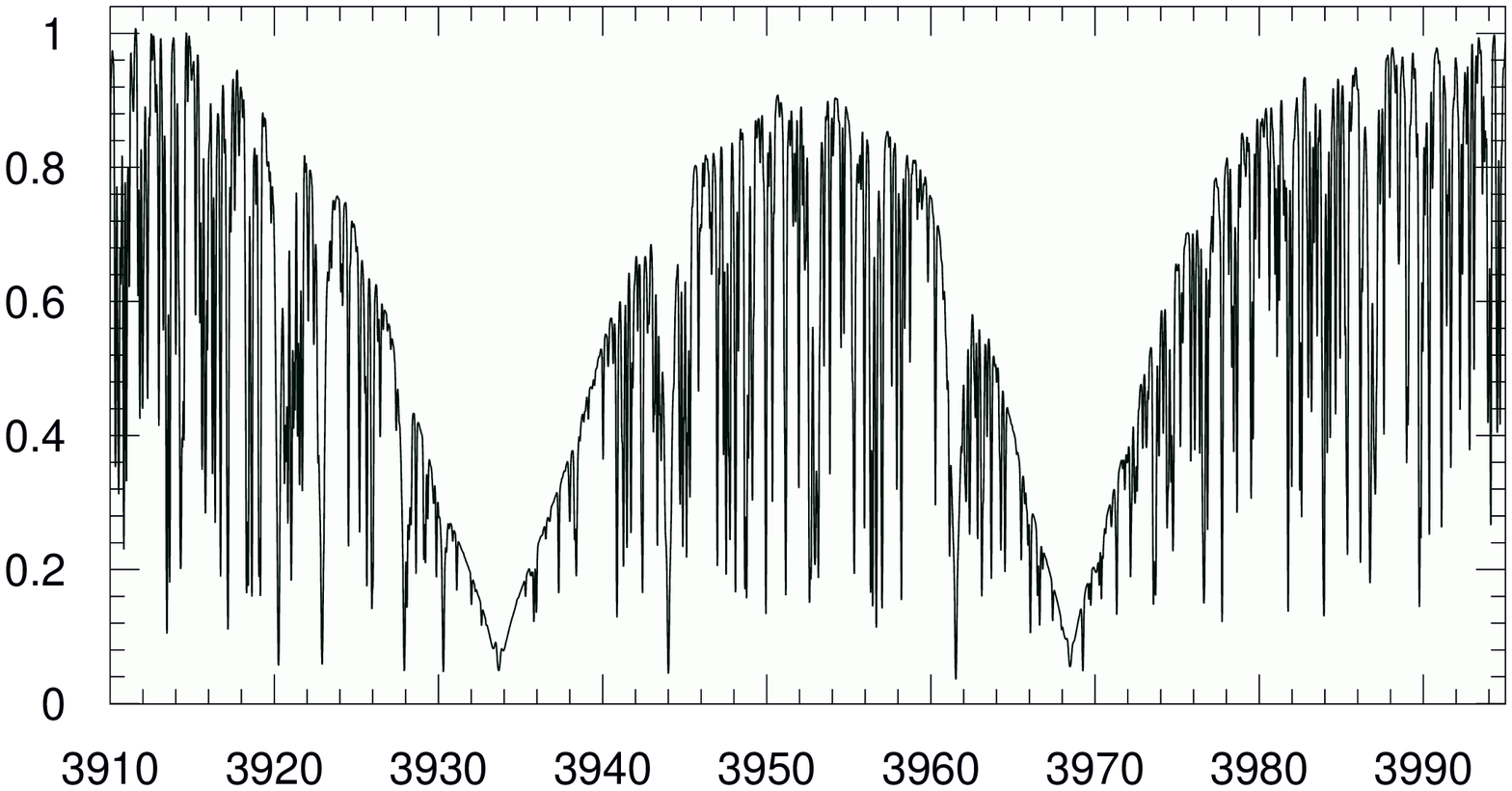}
{\bf b.}\\
\includegraphics[angle=0,width=87mm,clip]{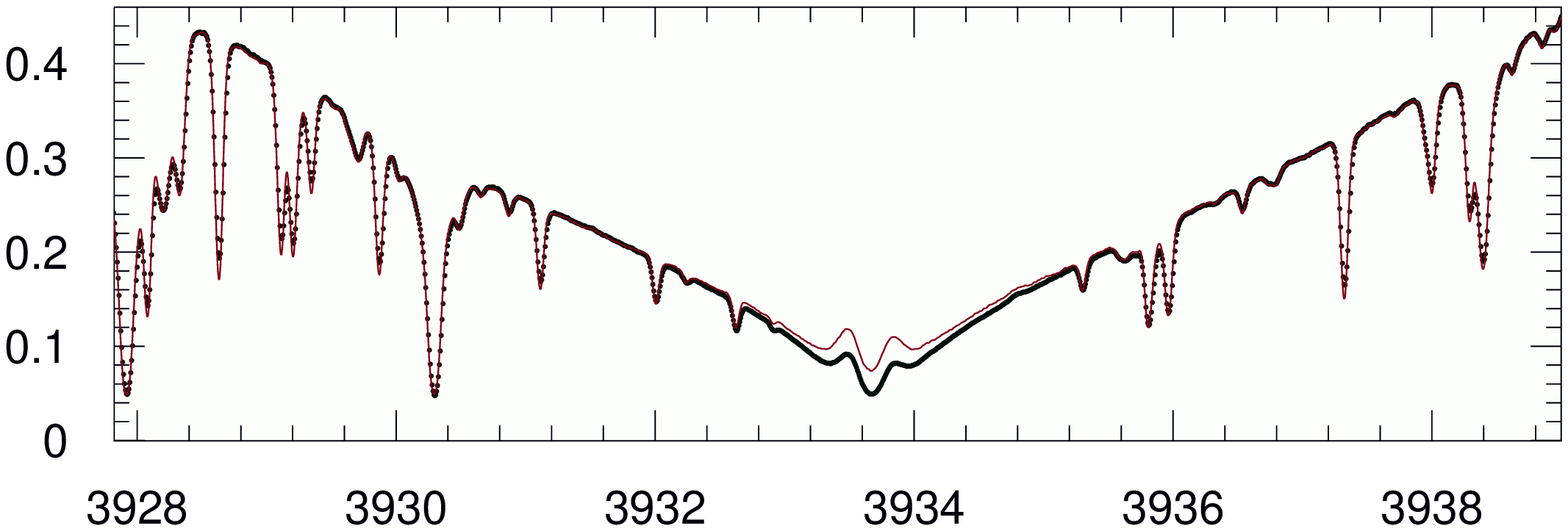}
{\bf c.}\\
\includegraphics[angle=0,width=87mm,clip]{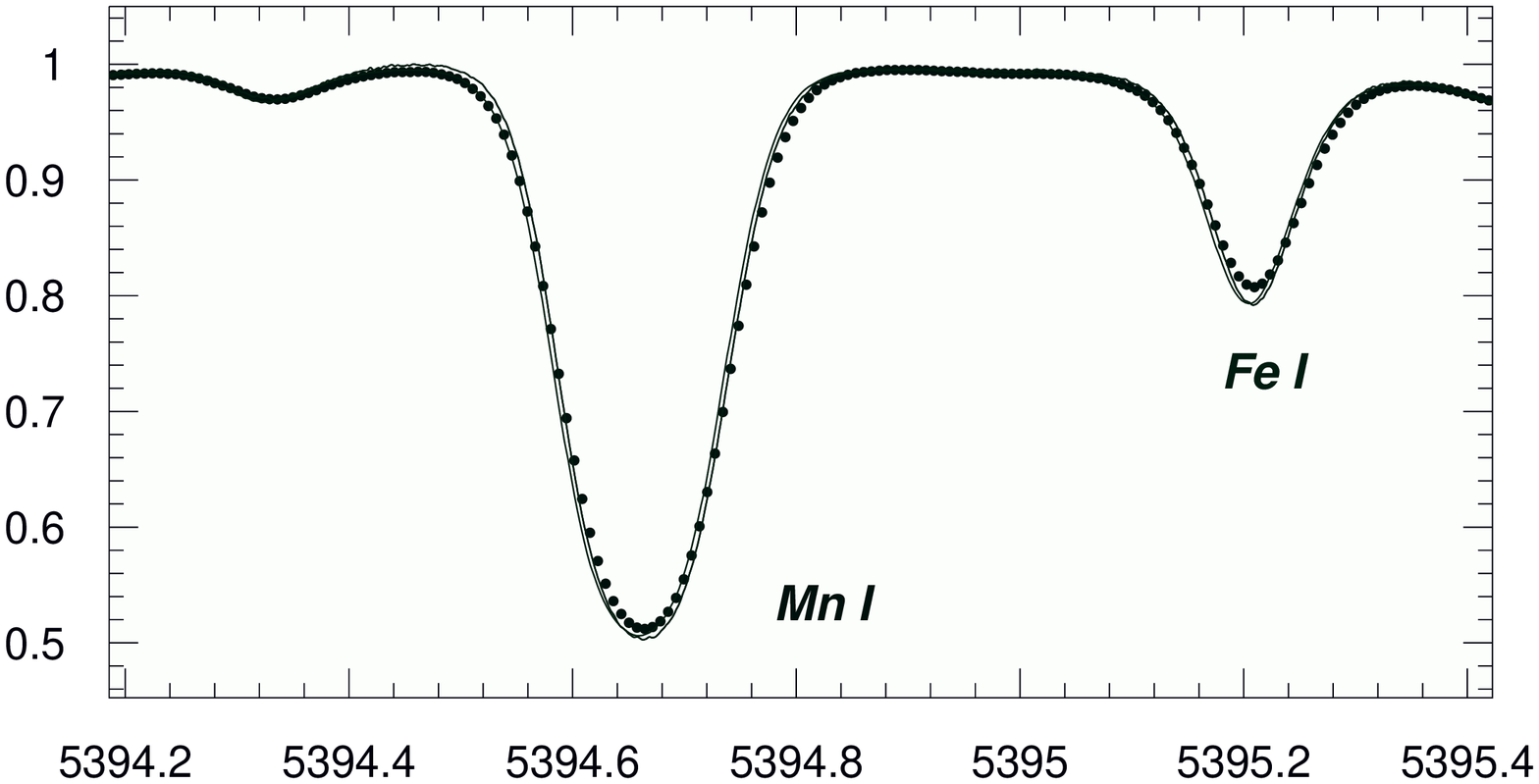}
{\bf d.}\\
\includegraphics[angle=0,width=87mm,clip]{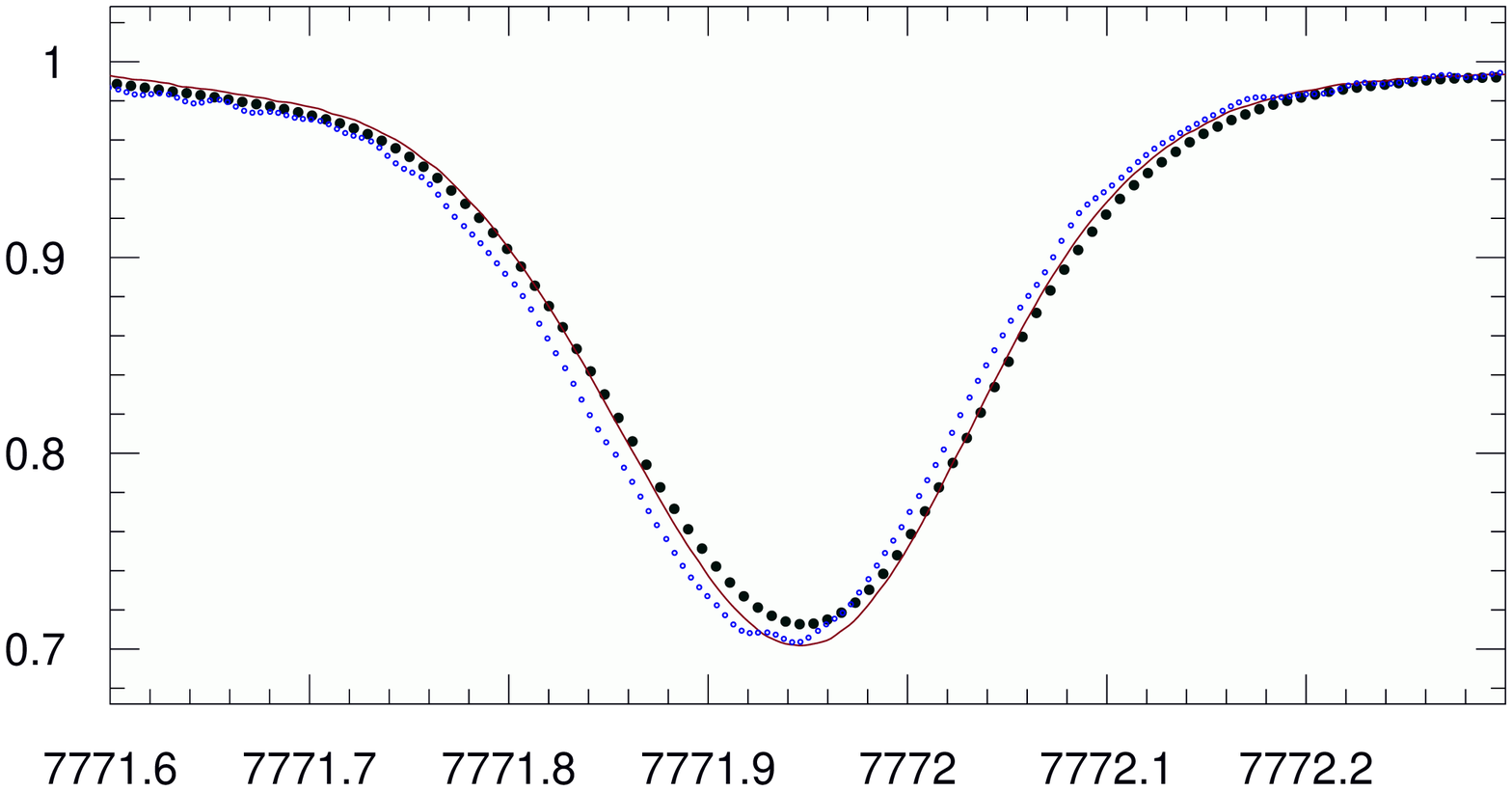}
\caption{Comparison of four representative wavelength portions with FTS solar atlases. The PEPSI spectrum is always plotted as big black dots except in panel $a$ where only the PEPSI spectrum is shown. Panel $a.$ shows the full Ca\,{\sc ii} H\&K doublet; $b.$ is a zoom into the core of the Ca\,{\sc ii} K line compared with the NSO FTS atlas (line); $c.$ shows the Mn\,{\sc i} 5394.7-\AA\ line. The NSO and IAG FTS atlases are plotted as lines (the more blueshifted line is the NSO atlas). Panel $d.$ shows the O\,{\sc i} triplet line at 7771.9\,\AA . In this panel the IAG FTS is plotted as a line and the NSO FTS as small blue dots. Both atlases are adopted with their nominal spectral resolution. }
 \label{F5}
\end{figure}

\section{Characteristics of the SDI spectra}\label{S4}

\subsection{Spectral resolution}

The effective spectral resolution $\lambda/\Delta\lambda$ for the data in this paper is obtained from the  full-width-at-half-maximum (FWHM) of the FPE fringes with the focus position 2990 and 7480 for the blue and the red arm, respectively. Focus sequences for all cross dispersers were made with a halogen lamp as the light source. An automated FFT auto-correlation of the approximately 1,700 fringes across the full CCD determines the FPE FWHM at any position on the CCD. In dispersion direction one \'echelle order is covered by a little more than 100 fringes (one fringe every $\approx$0.5\,\AA ). The FPE FWHM is then related to the Th-Ar line widths nearest in position on the CCD but from separate Th-Ar images. The intrinsic line width of the Th-Ar lines is taken from the NIST Th-Ar atlas (Kerber et al. \cite{nist:thar}) and calibrated against our Th-Ar widths and then removed from the PEPSI Th-Ar spectra. In this way, we determine the spectrograph focus at any location of the CCD and convert it to effective spectral resolution $\lambda/\Delta\lambda$. For the data in this paper the resolution ranges from an average of 220,000 in the blue to an average of 250,000 in the red. The respective deviation along an \'echelle order can be as large as $\pm$30,000, in particular for the red CDs. Note that these numbers are for the spectra from September 2015 and October 2016 with image slicer \#2, but will be closer to the nominal two-pixel resolution for future spectra because of a new image slicer implemented 12/2016.

The spectrograph resolution is shown in Fig.~\ref{F3}a versus wavelength for all CDs. Resolution varies along an \'echelle order by up to 30\%\ due to finite image quality, pixel under-sampling, and uncorrected \'echelle anamorphic magnification (cf. Schweizer \cite{anamorph}). This plot is deduced from a two-pixel sampling of the resolution element and represents the best-focus situation. In practice the spectral resolution depends on the achieved focus for the two optical cameras. There is residual astigmatism which could not be removed during the final alignment. If the focus achieved is, e.g., 2.5~pix FWHM, then the $R$ numbers in Fig.~\ref{F3}a must be multiplied by $2/2.5=0.8$. We note that the spectra in this paper were all taken with camera foci selected to suit all respective CDs simultaneously, i.e., without refocusing for particular wavelengths. One could refocus for every CD combination and reach a slightly higher average resolution but that was not done during the commissioning process leading to the present data. Therefore, resolution reaches 270,000 only at certain wavelengths in the yellow and red parts of the present spectra.

\subsection{Signal-to-noise ratio}

Table~\ref{T1} lists the measured S/N of the individual spectra and the S/N of the combined (``deep'') spectra at the central wavelength of each CD. S/N is always given per pixel. We see that the S/N for an individual integration runs from at worst $\approx$200 for the blue wavelength limit to up to a peak of $\approx$850 near 700\,nm and a fall off to $\approx$650 at the long wavelength limit. Daily deep spectra are built by average combining between 20 to 110 consecutive exposures, which boosts the S/N by factors of between $\approx$4--10. At the typical plotting scale from 0 to 1 in Fig.~\ref{F2} the (photon) noise in the spectra is not visible anymore (the noise seen in the continuum is from the over plotted HARPS spectrum). Fig.~\ref{F3}b is an example of the S/N of a single individual spectrum for the full PEPSI wavelength range. Normally, we record two SDI exposures simultaneously with two different fibres mimicking the two LBT sides. Their combination  further increases the S/N by a factor of $\sqrt{2}$ with respect to the numbers given in Table~\ref{T1}.

While the individual exposures are considered surface snapshots with respect to the five-minute oscillation pattern of the Sun, the deep spectra can be considered smeared surface averages and thus are better representatives of the average undisturbed photosphere. Their time coverage per CD ranges between 37\,min for CD\,I/404\,nm to 24\,min for CD\,VI/825\,nm.  Taking 20 individual exposures per CD thus samples the fundamental oscillation between 7.3 and 4.8 times from the bluest to the reddest wavelengths. The rms of these averaged spectra is not only expected to be a function of wavelength but also to vary with line strength in relation with the amplitude of the pulsation mode at that particular line-formation depth. Fig.~\ref{F-6428} shows a deep spectrum along with the individual spectra for a narrow wavelength strip of 1\,\AA\ centered around a weak line (Cr\,{\sc i} 6428.1\,\AA , equivalent width 1.8\,m\AA ). The un-weighted rms from these 20 spectra is less than 0.1\%/pix, just as expected from the individual S/N ratio. A variance spectrum and a periodogram also did not show a clear systematic pattern or signal. We conclude that we did not see intrinsic variability at these wavelengths for the duration of the exposure time at the date of our observations.

We note that for some wavelength regions the spectacular S/N of the average-combined spectra of several thousands to one is artificial because of the dominance of the fixed-pattern noise in some sections of the STA CCDs. These regions are limited to an equivalent S/N of approximately 1,300:1, independent of its exposure level (see Sect.~\ref{S4.4}). Unnoticed intrinsic solar changes in certain spectral lines during integration can further lower the actual S/N. For example, in the Ca\,{\sc ii} H\&K line cores, such changes dominate over photon noise already at a S/N of as low as 20:1.

\subsection{Wavelength scale and RV zero point}

Each individual solar spectrum is wavelength calibrated before average-combining them. The rms from $\approx$2,000 Th-Ar based wavelength solutions was 4--8\,\ms\ over the course of two weeks in September 2015 (depending on wavelength and always excluding CD-I). However, many individual spectra during this time showed more than twice this RV uncertainty. This was because the earlier version of the (robotic) SDI telescope introduced systematic guiding jitter due to disk vignetting in its fibre injection, a limitation also noticed by other groups (Dumusque et al.~\cite{dum}, Lemke \& Reiners \cite{lem:rei}). The new version of the SDI telescope (operative since October 2016) employs an integration sphere and a single off-axis reflecting telescope and is now essentially free of guiding jitter. The spectra in this paper have an average rms from a series of Th-Ar solutions of 1.2\,\ms\ (see later Sect.~\ref{S51}). The remaining uncertainties are the sum of several blemishes described in the next chapter. For example, gradients of the CCD temperature (from one corner of the chip to another) can be as much as 18\,mK and cause RV variations of up to 100\,\ms . However, these are easily identified and corrected due to their persistent nature and were removed for the RV rms measurements above. Cross correlations of PEPSI spectra in the 450--600\,nm region with the laser-comb calibrated HARPS solar spectrum from Molaro et al. (\cite{molaro}) revealed an average RV deviation of just 10\,\ms .

We recall that the wafer mask of the STA1600 CCDs is created by raster scanning the desired pattern and then imprinting on the wafer using an electron beam or laser. Only a small section of the mask is written initially, most of the wafer-sized CCD area is created by mechanically moving the small section to bring the next area under the beam. Small errors in the mechanical steps of the mask motion can show up as a pattern that is effectively a pixel-size variation on the CCD at the borders between consecutive mask steps. This problem had been identified by STA who created now higher resolution masks for better uniformity (Bredthauer \& Boggs \cite{bb}). However, our current CCDs were still produced with the older and lower-resolution masks.

Similar effects arise from a stepper mask CCD production, for example in 512-pixel sections depending on the producer. Its stitching introduces systematic pixel-size discontinuities. For the HARPS e2v CCDs this resulted in distortions of the Th-Ar wavelength solution of $\pm$40\,\ms\ along an \'echelle order and across the entire wavelength range (Molaro et al.~\cite{molaro}). The stable and accurate zero point from laser combs pinpoints these CCD deficiencies and allows us to correct for them. No such stitching of subsections exists for the PEPSI STA CCDs due to the different production scheme but the mask motion errors introduce in practice the same problem (see Sect.~\ref{S3.3}).

\subsection{Blemishes}\label{S4.4}

\emph{Residual CCD fixed-pattern noise}. There are several sources that contribute to fixed-pattern noise in our CCDs. Firstly, the PEPSI STA CCDs have 16 amplifiers which are grouped in two rows with eight amplifiers on adjacent sides of the chip. This means that every \'echelle order once crosses an amplifier border in dispersion direction (in the middle of the chip). In cross-dispersion direction the CCD is sectioned eight times. Because every amplifier section is read out with its own gain factor, there is at least one intensity jump along the dispersion trace and up to seven jumps along a cross-dispersion trace. The flat field division takes care of these jumps unless there were unnoticed gain variations. No inter-amplifier space exists and therefore does not interrupt the wavelength solution. Secondly, we noticed a periodic fixed-pattern noise in the master flat fields at very high exposure level that results likely from a back reflection from the CCD substrate. This effect is not visible in normal target exposures but can be noted in single flat-field exposures if exposed near the edge of the linearity regime. The simple cure of the problem is to use flat-field exposures well below the full well of the chip with count rates similar to the target count rates. The practical problem is that one needs at least 30 well-exposed flat fields to make up a master flat appropriate for the high-level average-combined solar exposures.

Thirdly, pixel photo response nonuniformity causes random and structured patterns from errors in the (CCD)mask-creation process. Not all of the response nonuniformity patterns are perfectly removed by the regular master-flat division because the CCD response appears to be a function of ADU counts in some sections (``taps'') in our devices (a tap relates to an amplifier). For those pixels, we find that the gain varies by approximately 5\,\% (from zero ADU to saturation ADU) and also appears with a pattern with a main spacing of 38 pixels along rows. Thus, the response of some of the 112 million pixels is nonlinearly increasing with exposure level. We incorporate this behavior in the standard data reduction process by using a ``super master flat''. Such a super master flat takes about 30 master flats with different illumination levels ranging from nearly zero to saturation. One master flat is already the sum of 70 regular flats (per CD setting) and thus our super master flat consists of approximately 2,000  individual flat-field exposures. The remaining (residual) pixel-to-pixel nonuniformity after the super master flat division currently limits the S/N of the affected CCD taps to a conservative $\approx$1,300:1. It leaves ripples with a main periodicity of 38 pixels in the spectrum. While the ripple's amplitudes are larger on the blue CCD, where basically no wavelength dependency is seen, the amplitude is only half of that on the red CCD but a wavelength dependency exists. A polynomial fit for every exposure level of every pixel for every CD setting is then tabulated and used in the automatic data reduction.

\emph{Order merging}. The spectrum at the beginning and the end of an \'echelle order repeats in adjacent orders. To average the over-lapping parts of the orders, one must merge spectra from opposite sections of the CCD. The two-dimensional pixel-to-pixel sensitivity is taken out by the flat-field division while we minimize its residuals in time by an illumination-dependent super master flat field (see above). But the remaining issue in PEPSI is the non-telecentric focus deviation across the geometric field of view of the huge CCDs (10\,cm$\times$10\,cm). It causes the focus and thus the Th-Ar line width to be different at opposite edges of the CCD. Fig.~\ref{F3}a demonstrates this in terms of the FWHM of Th-Ar emission lines. We proceed with order merging by re-sampling the overlapping orders to an average dispersion before co-adding with a weight function which is the inverse of the respective pixel variances. One could dub this procedure optimal merging. Nevertheless, it could introduce a wavy continuum rectification around the merging wavelengths. The present spectra are free of such an effect above the 0.1\%\ level.

\emph{Partial order extraction}. Due to the tilt of the dispersion direction with respect to the CCD rows, the very first \'echelle order (\#160 in CD-I) and the very last order (\#68 in CD-VI) are recorded only partially (see Fig.~15 in Strassmeier et al.~\cite{pepsi}). These two orders cover the extreme wavelengths of the spectrograph at wavelengths longer than $\lambda$9001.6\,\AA\ and shorter than $\lambda$3849.8\,\AA . Its extraction and rectification suffers greater uncertainty than the fully recorded orders, suggesting equivalent-width measurements of these regions to be less precise. Note that the orders on the edge of the other CDs are also just partially recorded but are not used because these orders repeat as full orders in adjacent CDs.

\subsection{Comparisons with other solar atlases}

Fig.~\ref{F5} shows several representative comparisons with the Kitt Peak NSO FTS flux atlas (Wallace et al.~\cite{nso}) and the new IAG FTS atlas (Reiners et al.~\cite{iag}). The wavelength examples show the full Ca\,{\sc ii} H\&K doublet, a zoom into the core of the K line, the Mn\,{\sc i} 5394.7-\AA\ line, and one of the O\,{\sc i} triplet lines at 7771.9\,\AA . The Ca\,{\sc ii}~K line core (Fig.~\ref{F5}b) shows a markable difference between the NSO FTS atlas taken in 1981 near the peak of solar cycle 21 and the PEPSI spectrum (dots) taken in 2016 in the middle of the declining activity of cycle~24. Cycle~24 was already among the weakest cycles ever recorded in history (e.g., Jiang et al. \cite{jcs}) and our H\&K spectrum seems to reflect this. Note that the IAG FTS does not cover the 390\,nm region and thus allows no comparison. A total of three such deep spectra of the full PEPSI wavelength range, and with the time coverage indicated in Table~\ref{T1}, are provided in this paper for further comparisons.

Our spectra fit in resolution in between the FTS flux atlases and the HARPS atlas. The wavelength range in Fig.~\ref{F2} shows several line cores of relatively strong lines where this is obvious. For most weaker lines, e.g. C\,{\sc i} in this figure, the FTS and the PEPSI spectra are basically indistinguishable while the HARPS spectrum appears noticeably weaker in the line core and more noisy. At this point, we note again that the PEPSI solar spectra in this paper did not reach the full 2-pixel resolution at all wavelengths but about 80\%\ of it on average. For comparison, lower than predicted resolution was also found for the Li\'ege disk-center atlas by Doerr et al. (\cite{doerr}), but with even two to six times lesser resolution than specified.

\begin{figure}
\includegraphics[angle=0,width=87mm]{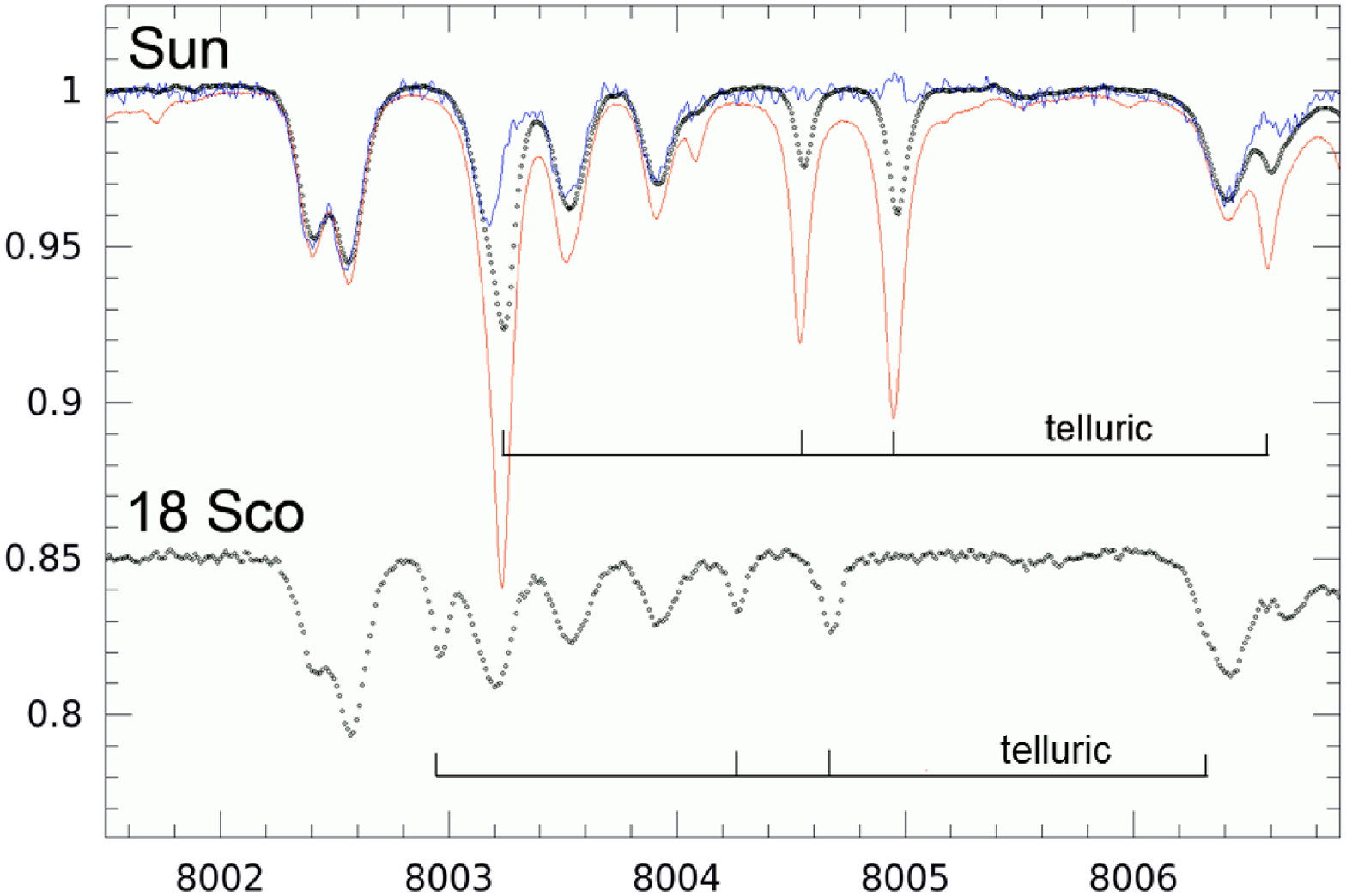}
\caption{Terrestrial water-vapor lines in the CN-line region near 8000\,\AA\ for the Sun (top) and the solar-twin 18\,Sco (bottom; shifted by --0.15). The PEPSI spectra are shown as dots. The solar spectrum is compared with the IAG FTS atlas (the deep-lined spectrum in red) and the ``dry'' NSO FTS atlas (weaker lined in blue). The IAG FTS spectrum is contaminated by very strong telluric lines while the NSO spectrum is comparably more noisier. Note that the 18~Sco spectrum was shifted to match the solar spectrum.}
 \label{F-8003}
\end{figure}

The continuum levels for wavelengths longer than $\approx$460\,nm agree generally very well with the FTS atlases except maybe at regions contaminated by telluric bands. Several strips of spectrum with almost no lines indicate offsets of 0.7\%\ on average. For wavelengths below $\approx$460\,nm the differences are only marginally larger until $\approx$390\,nm, say 1\%. However, for wavelengths shorter than this the discrepancy gets increasingly worse, reaching 10\%\ or more at 384\,nm. This is mostly due to limited S/N, unaccounted stray light, and the fact that there is no continuum seen because of the overwhelming line blanketing. For wavelengths shorter than 390\,nm, we consider the NSO FTS atlas more reliable than PEPSI.

The LBT site on Mt. Graham in Arizona can be a very dry place and accordingly weak is the intrinsic telluric water contamination in the PEPSI spectra. Of course, this depends on the observing season. The considerable  atmospheric contamination of the solar spectrum led Wallace et al. (\cite{nso}) early on to observe the Sun at various air masses and then remove the telluric contamination. Their spectrum represents a ``dry'' optical solar flux spectrum. A close comparison of a wavelength region of general astrophysical interest that is affected by telluric lines is shown in Fig.~\ref{F-8003}. This wavelength region contains several $^{12}$CN and $^{13}$CN blends that are commonly used to determine the carbon isotope ratio $^{12}$C/$^{13}$C, which is a primary indicator for the evolutionary status of the stellar fusion process. The figure also shows a PEPSI spectrum of the solar-twin 18\,Sco for comparison with the Sun. Its telluric contamination appears shifted by $-0.26$\,\AA\ with respect to the solar spectrum (because of the difference of radial velocity) but can be otherwise directly compared with the solar spectrum. It is immediately seen that the IAG FTS atlas suffers from very strong telluric lines due to being taken in the city of G\"ottingen in Germany and that the NSO FTS spectra are approximately three times noisier in this wavelength region than even the LBT/PEPSI 18\,Sco observation. All spectra in Fig.~\ref{F-8003} are plotted at their respective spectral resolution though.

\begin{figure}
{\bf a.}\\
\includegraphics[angle=0,width=87mm]{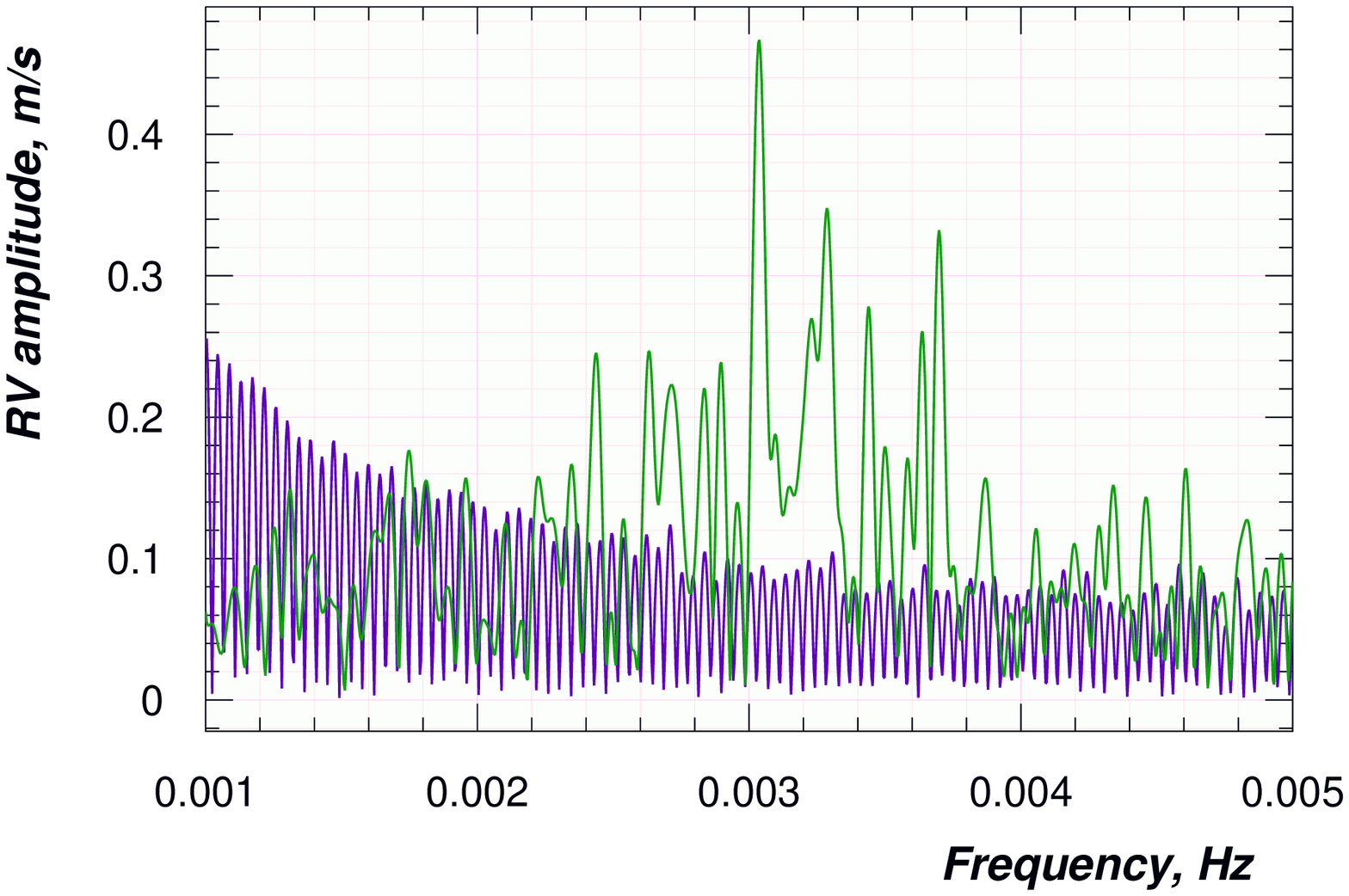}
{\bf b.}\\
\includegraphics[angle=0,width=87mm]{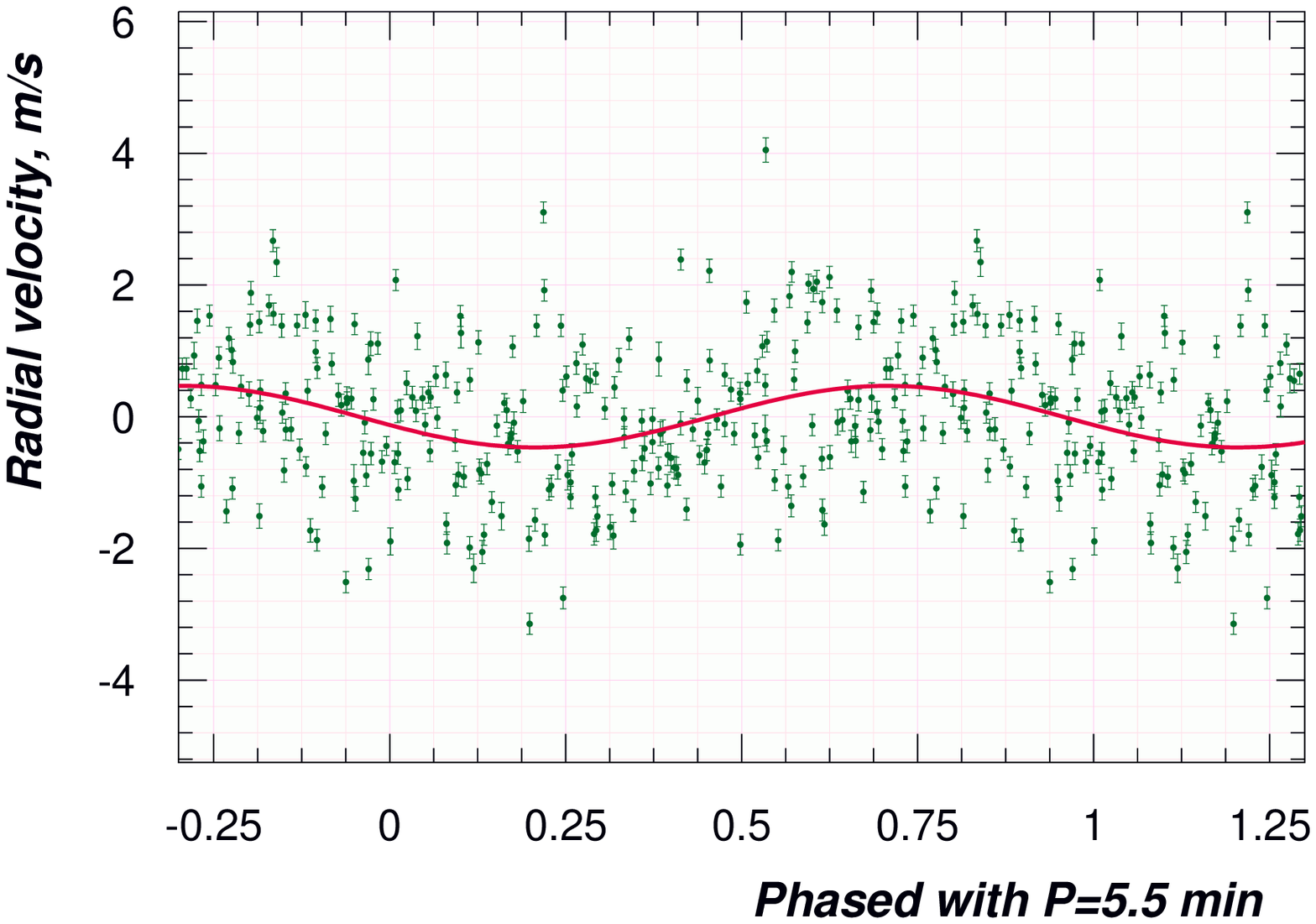}
\caption{Detection of solar p-mode oscillations with Sun-as-a-star PEPSI spectra. $a.$ Frequency fits  from the time series of 8.5 hours on Oct.~15, 2016. The peak at 3\,mHz (5.5\,min) appears with an amplitude of 47\,cm\,s$^{-1}$ (the blue line is the window function). $b.$ RV versus phase combined with the best-fit period. Its rms is 1.2\,\ms . }
 \label{F-pmode}
\end{figure}

\section{Science examples}\label{S5}

Many archival science cases could be tackled with our spectra, e.g., the many spectral lines may be measured for systematic RV changes to determine convective blueshift (Dravins et al. \cite{dravins}, Reiners et al. \cite{iag}), or elemental abundances of some rare-Earth elements having only extremely weak lines, or the iron mass fraction through Fe/Si abundance ratios that has been proposed to be a critical ingredient for rocky planet models (Santos et al. \cite{santos}). For this initial paper, we pick out two early-science examples.

\subsection{Solar p-mode oscillation from disk-averaged spectra}\label{S51}

Many physical processes and phenomena contribute to the solar disk RV profile (see, e.g., Pall\'e et al. \cite{palle95}, Meunier et al. \cite{meun}, Lanza et al. \cite{lanza}, Haywood et al. \cite{hay:acc}). Besides global phenomena like rotation and differential rotation, local phenomena like spots and plages and their associated local velocity fields significantly contribute to the disk-averaged radial velocity at any given time, and also vary with the solar cycle. The famous acoustic oscillations (p-modes) are even acting on various length scales, from global to local, and additionally also on short (minutes) and medium (days) time scales. Due to the stochastic-excitation nature of p-modes disk-integrated spectra taken over longer periods of time are difficult to combine; as nicely demonstrated by the SOHO/GOLF data (e.g., Pall\'e et al. \cite{palle99}). Most of the oscillatory power is between 2.5-4.5\,mHz and a length scale of close to 10$^4$\,km. However, a short-term high-cadence sequence should preserve the phase of the typical 5-min oscillation even from (disk-averaged) radial velocities. Such an observing sequence was obtained with SDI and PEPSI for demonstration purposes on three consecutive days during October 14-16, 2016.

A total of 996 back-to-back spectra were taken with CD\,III ($\lambda$480--544\,nm) and resulted in  approximately 300 velocities every day for 8.5~hours. Exposure time was 3\,s. Air mass varied between 1.4 to 4.4 and S/N between 490 to 410, respectively. Figure~\ref{F-pmode}a shows the results from the second day (Oct.~15, 2016). The two other days are consistent with this result. However, the combination of all three days results in a phase smearing which diminishes the RV amplitude below our detection threshold, as is expected from the stochastic excitation of the oscillations and its coverage with a cadence of 92\,s.

\begin{figure}
{\bf a.}\\

\includegraphics[angle=0,width=87mm]{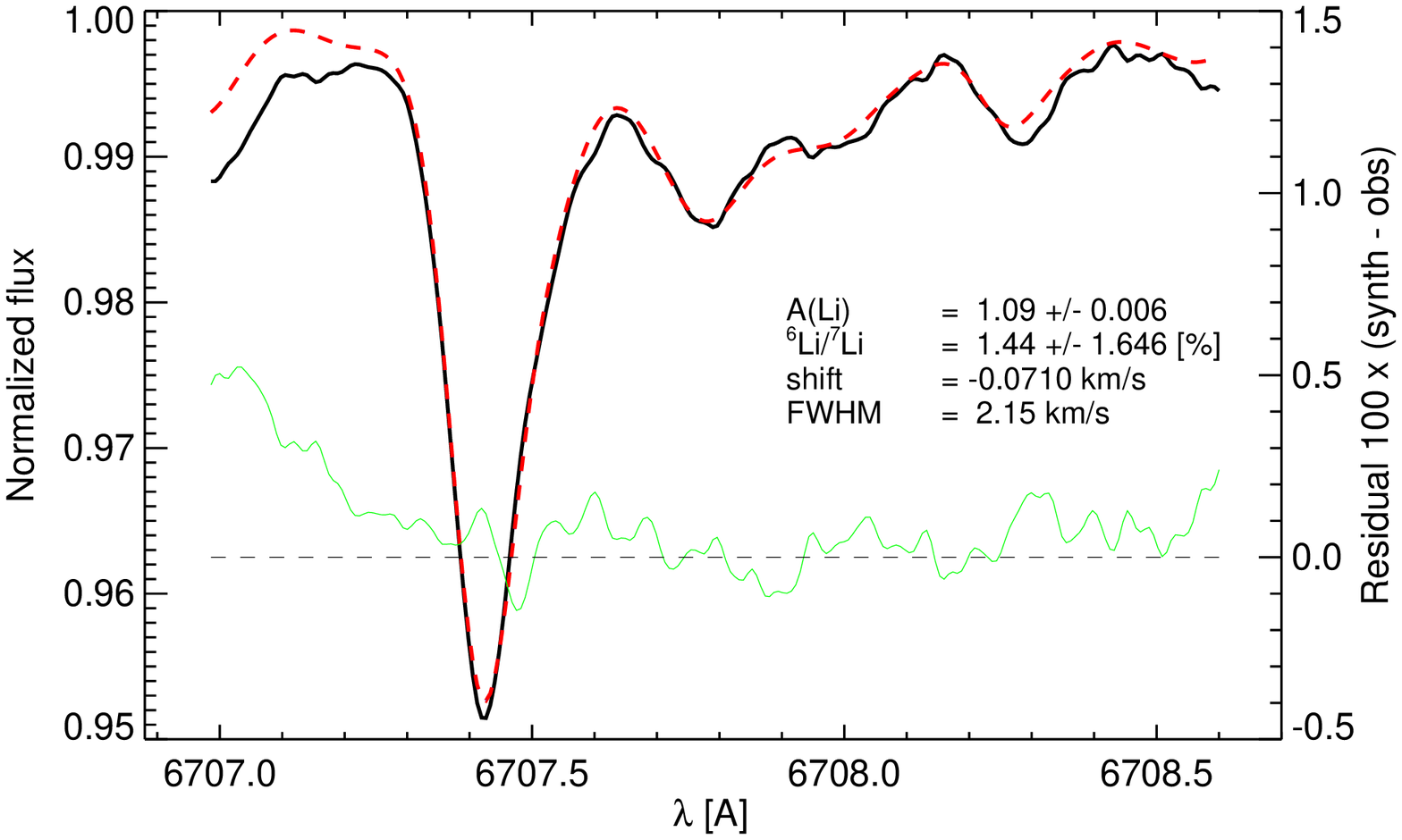}
{\bf b.}\\
\includegraphics[angle=0,width=87mm]{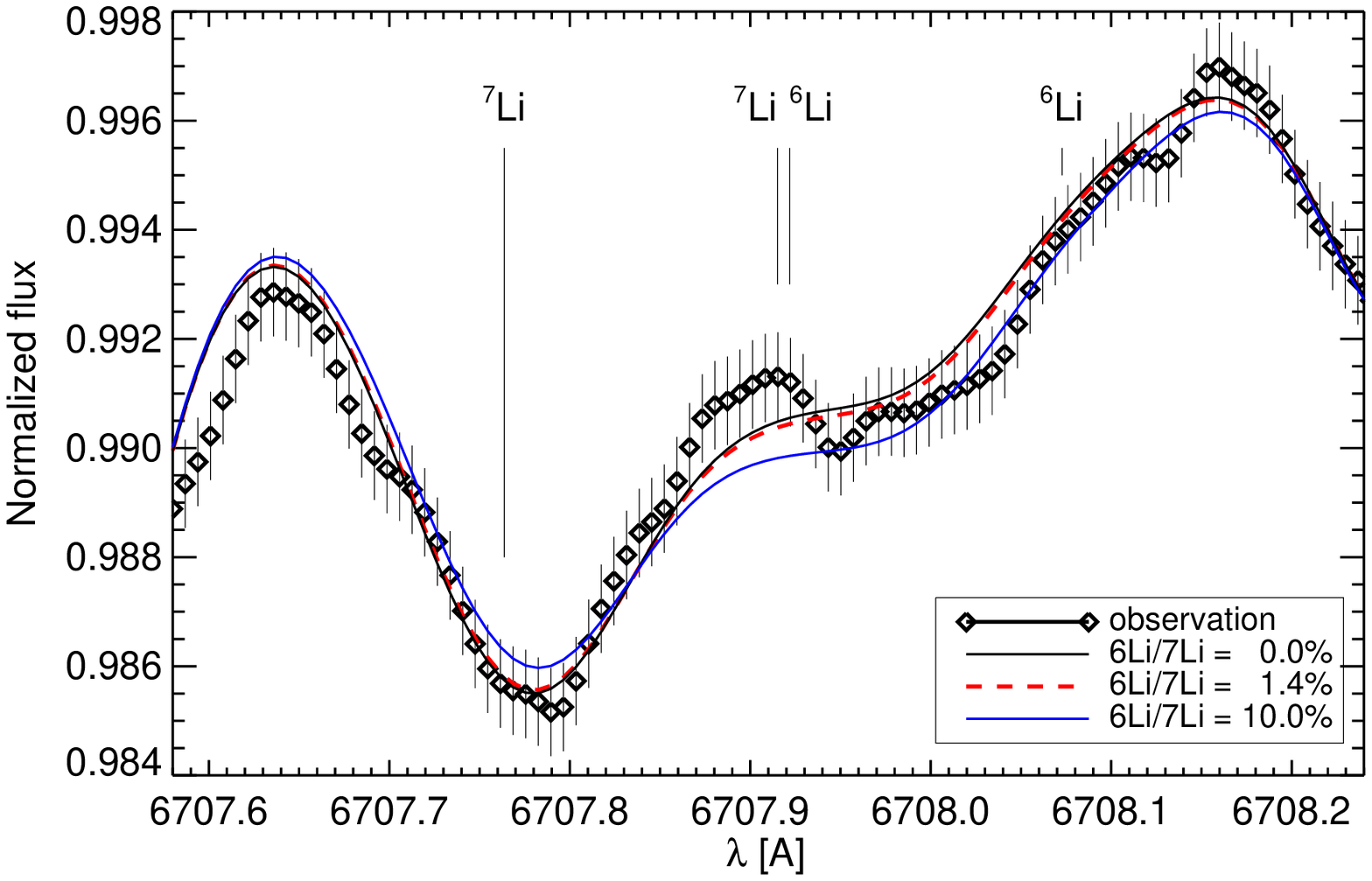}
\caption{Li\,{\sc i} 6707.8-\AA\ line of the Sun. \emph{a.} A 1.8-\AA\ section of a deep PEPSI spectrum (thick black line) and the fit with a synthetic 3D NLTE spectrum (dashed line). The insert lists the best-fit free parameters. The residuals (thin lines) are enhanced by a factor 100 for better visibility and offset to $\approx$0.96 units in flux. \emph{b.} A zoom into a 0.6\,\AA\ subsection centered on the two Li doublets. Observations are now indicated as open diamonds with error bars. The colored lines are based on the best fit from panel $a$ but for three different isotope ratios as indicated in the insert. The dashed line is the best fit.}
 \label{F-Li}
\end{figure}

For the period search we performed a least-squares fit for each trial frequency with subsequent statistical F-test at each step against the null hypothesis of the absence of periodicity. For all three days we consistently found a preferred frequency at 3.04\,mHz, corresponding to a period of 5.5\,min. It peaks at an amplitude of 47\,cm\,s$^{-1}$ with a significance of 5.2$\sigma$  with a FAP of 5$\times$10$^{-4}$ and $\chi^2=106$. The extended scatter of the data around the best fit is caused both by the seeing effect due to short integrations as well as the super-granulation variations on the surface of the Sun. The peak with the highest significance is used for the non-linear least-squares fit to the data (Fig.\ref{F-pmode}b). The daily rms RV scatter is 1.2\,\ms .

\subsection{Solar Li abundance}\label{S52}

The analysis of isotopic ratios and their relative contributions requires extremely clean, high-quality spectra. We recall that the equivalent width of the Li\,{\sc i} doublet is typically 1~m\AA\ in the interstellar medium. Disk-resolved solar observations show not only significant center-to-limb variations of the many CN blends that affect the Li line but also a larger Li equivalent width at the limb than in the disk center (3.7 vs. 5.7\,m\AA ; Brault \& M\"uller \cite{bra:mue}). M\"uller et al. (\cite{mue:pey}) showed this to be partially a non-LTE effect and that the elemental abundance is actually the same. Their analysis resulted in $A(\mathrm{Li})$=1.0$\pm$0.1 and $^6$Li/$^7$Li$\leq$0.01; our today's still accepted solar lithium abundance and isotopic ratio\footnote{throughout this paper, we give logarithmic abundances, $A(\mathrm{Li})$, on a scale defined by $A(\mathrm{Li}) = \log n(\mathrm{Li}) - \log n(\mathrm{H}) + 12.00$}. A Li determination from a high-resolution spectrum of a sunspot umbra by Ritzenhoff et al. (\cite{rit:sch}) resulted in a LTE abundance of 1.02$\pm$0.12 together with an upper limit for the isotope ratio of 0.03. Despite that its analysis lacked a full non-LTE treatment and that additional uncertainty is involved from the not well-known TiO contamination in cool sunspots, it indicated the same Li abundance than for the undisturbed photosphere. We also note that the $\leq$0.03 isotope ratio is in agreement with the solar-wind Li isotope ratio of 0.032$\pm$0.004 measured in lunar soil (Chaussidon \& Robert \cite{cha:rob}).

More recently, Caffau et al. (\cite{caf:sun}) compared synthetic 3D NLTE spectra with the solar spectra of Neckel \& Labs (\cite{nek:lab}) and of Kurucz (\cite{kur}) but assumed no $^6$Li contribution. Their final choice for the blending line-list was that of Ghezzi et al. (\cite{ghezzi}) with a few updated oscillator strengths. A logarithmic abundance of 1.03$\pm$0.03 was obtained, well within the error limits of the M\"uller et al. (\cite{mue:pey}) and Ritzenhoff et al. (\cite{rit:sch}) results, where the uncertainty reflects the dispersion among the results for the different observed spectra rather than deficiencies in the line list or the model. The photospheric solar lithium abundance is thus approximately 160 times lower than that measured in meteorites (Asplund et al. \cite{asplund}). The isotopic line ratio in the interstellar medium was determined earlier to $\approx$8\%\ by Lemoine et al. (\cite{lem:fer}) and others, and matches well the meteoritic value of 8.33\%\ listed in the review by Asplund et al. (\cite{asplund}).

We apply the most recent 3D hydrodynamical CO5BOLD model atmosphere of the Sun (Ludwig et al. \cite{cifist}). A full NLTE line synthesis is done with NLTE3D+LINFOR3D\footnote{http://www.aip.de/Members/msteffen/linfor3d} in combination with the recent blending line list from Mel\'endez et al. (\cite{mel:ber}) as well as Ghezzi et al. (\cite{ghezzi}). Our synthesis also considers the full isotopic and hyper-fine structure represented by twelve components, six of which belong to $^6$Li and six to $^7$Li (a table of all blending transitions is available in Caffau et al. \cite{caf:mot} and Mott et al. \cite{mott}). We also adjusted the transition probability of a vanadium blend (Lawler et al. \cite{lawler}; see Mott et al. \cite{mott} for application details).

We started by comparing the 670.70--670.87\,nm region of all spectra listed in Table~\ref{T1} and inspecting them for systematic differences like unnoticed small cosmic-ray hits or residual continuum offsets and wavelength shifts. Then we build three daily deep spectra out of the pool in Table~\ref{T1} and use them for the analysis. These data are not limited by photon noise anymore but by the residual fixed-pattern noise of the CCD (see Sect.~\ref{S4.4}). Unfortunately, the Li wavelength region is affected by this. Fig.~\ref{F-Li}a shows a 1.8-\AA\ section of one deep spectrum with a nominal S/N of 8,000:1 which is, however, reduced at the Li wavelengths to 1,300:1. We employ the least-squares fitting algorithm MPFIT (Markwardt \cite{mpfit}) in a procedure described in detail in Steffen et al. (\cite{steff}). It adjusts iteratively four fitting parameters until the best fit (minimum $\chi^2$) is achieved. The four free parameters were $A(\mathrm{Li})$, $^6$Li/$^7$Li, a global wavelength shift, and a global Gaussian line broadening (FWHM). Rotational broadening $v\sin i$ of 1.8\,\kms\ and the level of the continuum were kept fixed (the instrumental broadening at the Li wavelength is 1.37~\kms ). The best fits for the Li lines are always achieved with the line list of Ghezzi et al. (\cite{ghezzi}) expanded with the revised $\log gf$ for V\,{\sc i} 6708.1\,\AA\ from Lawler et al. (\cite{lawler}). We also increased the $gf$ values of all CN lines by 30\%\ because Ghezzi et al. had used a higher solar CNO abundance than we do in this paper, and this compensates for it. The formal $\chi^2$ fit level is 966 compared to 388 with the expanded Mel\'endez et al. (\cite{mel:ber}) list. However, the formal $\chi^2$ is not relevant in our case because it includes the full spectral range where the Mel\'endez et al. line list better fits the edges of the spectral range than Li, in particular the Fe\,{\sc i}+CN blend. Considering only the Li-line range, the Ghezzi et al. line list is a better fit. Both line lists converge on an identical Li abundance though. Our best-fit value is $A(\mathrm{Li})$=1.09$\pm$0.006$\pm$0.04. The $\pm$0.006 is the internal fit error, the $\pm$0.04 is estimated from repeated measurements with different continuum locations, the different line lists, and the three different observations. The best-fit isotope ratio $^6$Li/$^7$Li is 1.4$\pm$1.6\,\%, and thus appears consistent with zero $^6$Li content.

\section{Summary and conclusions}\label{S6}

We present a homogeneous set of solar flux spectra taken with the new PEPSI spectrograph of the LBT and use it to verify the capabilities of the instrument. Detailed comparisons with solar FTS atlases, as well as with the HARPS solar atlas, validate the future PEPSI data product. We found excellent intensity agreement down to the sub-percent level and a spread of RVs with an rms of 10\,\ms\ with respect to the HARPS solar atlas. The photon noise in the continua of the deep spectra is boosted to a S/N of $\approx$8,000:1 for the red  wavelength regions, and still 600:1 for the Ca\,{\sc ii} H\&K line core. The continuum placement uncertainty is only sub-percent for wavelengths longer than $\approx$400\,nm but deviates up to 1\% at 390\,nm  and up to 10\%\ at the blue cut-off wavelength at 384\,nm. The latter mostly due to the inherent line blanketing at these wavelengths and the limited S/N ratio. We recall that the resolving power is not constant along the entire wavelength range of PEPSI, i.e. spanning 530\,nm, and also varies along an \'echelle order, itself on average $\approx$6\,nm wide. The nominal two-pixel resolution is 270,000$\pm$30,000 ($\langle$FWHM$\rangle$ of 1.11\,\kms ) but the data in this paper have typically 240,000$\pm$30,000 ($\langle$FWHM$\rangle$ of 1.25\,\kms ) due to imperfect camera focus (red and blue cameras) at the time of the observations.

Three mini time series of $\approx$300 spectra per day with a cadence of 92\,s are used to detect the disk-integrated p-mode oscillations of the Sun. We employed PEPSI's CD-III ($\lambda$480-544\,nm) only, and found a disk-averaged RV amplitude of 47\,cm\,s$^{-1}$ with a period of 5.5\,min. This value is within the measurements from SOHO/GOLF (Garcia et al. \cite{garcia}). GOLF measured disk-averaged RVs from the Na\,D doublet wings and found amplitudes between basically zero and up to 5~\ms\ depending on time. Our measurement is just a (daily) snapshot and more representative of the undisturbed solar photosphere because we used basically all detectable spectral lines within CD\,III. The one PEPSI measurement and the GOLF time series are hardly comparable because of the large time difference but also because the sodium D lines are affected by magnetic activity, but nevertheless agree with each other.

The deep spectra are used to re-determine the solar Li abundance. A 3D NLTE fit with the line list of Ghezzi et al. (\cite{ghezzi}), expanded by a vanadium blend and increased CN strength, provides the best-fit solution of $A(\mathrm{Li})$=1.09$\pm$0.04. An isotope ratio $^6$Li/$^7$Li of 1.4\,\%\ remained insignificant though and is consistent with zero $^6$Li. The derived Li abundance is fully consistent with previous estimates found in the literature. Grevesse \& Sauval (\cite{gre:sau}) and Lodders (\cite{lodd}) both quote $A(\mathrm{Li})$=1.10$\pm$0.10, while Asplund et al. (\cite{asplund}) recommend 1.05$\pm$0.10. A slightly lower value of 1.03$\pm$0.03 is given by Caffau et al. (\cite{caf:sun}). The solar system isotopic ratio is $^6$Li/$^7$Li$\approx$0.08 (Lodders \cite{lodd}). Because the lighter isotope $^6$Li is significantly more fragile than the main isotope $^7$Li, i.e. is more easily destroyed by nuclear reaction with protons, the depletion of $^6$Li in the solar convection zone must proceed at a much higher rate than that of $^7$Li. Hence it is expected that $^6$Li is completely destroyed in the solar convection zone. Any detection of $^6$Li in the solar photosphere would imply a very efficient external production mechanism. However, our measurement is consistent with zero photospheric $^6$Li content.

We conclude that the night-time spectrograph PEPSI can deliver high-quality spectra both in terms of intensity and radial velocity. No conclusions can be made on its long-term RV stability yet but data of the bright planet-host stars $\tau$\,Boo and 51\,Peg are being taken as a reference through the PEPSI-VATT fiber connection.

The reduced 1D deep spectra as well as the time series 1D spectra can be downloaded in FITS format from our web page at
\begin{center}
https://pepsi.aip.de .
\end{center}
On demand, we also provide the data prior to continuum rectification.

\acknowledgements{We thank all engineers and technicians involved in PEPSI, in particular our Forschungstechnik team and its late Emil Popow who passed away much too early. Special thanks are also due to LBTO's Christian Veillet, Mark Wagner and John Little and the entire LBTO mountain crew and particularly to Mike Lesser and his University of Arizona ITL CCD aficionados without whom we could not have solved the zillions of problems. Finally, we want to thank the referee for his/her careful reading which clearly improved the paper. It is also a pleasure to thank the German Federal Ministry (BMBF) for the year-long support for the construction of PEPSI through their Verbundforschung grants 05AL2BA1/3 and 05A08BAC that also were the basis for our involvement in the next generation ELT spectrographs. This research has made use of NASA's Astrophysics Data System and of CDS's Simbad database.}

\end{document}